\documentclass[preprint,aps,prl,showpacs,floatfix]{revtex4}
\usepackage{amsmath}
\usepackage{times}
\usepackage{euscript}
\usepackage{graphicx}
\usepackage{amsthm}
\usepackage{epsf}
\usepackage{epsfig}
\usepackage{bm}
\usepackage{soul}
\usepackage{xcolor}
\graphicspath{ {./figs/} }
\renewcommand{\vec}[1]{{\mbox{\boldmath$#1$}}}

\usepackage[linktocpage=true,plainpages=false,pdfpagelabels=false]{hyperref}
\begin{document}
\thispagestyle{empty}
\title{How to observe the vacuum decay in low-energy heavy-ion collisions}
\author{I.~A.~Maltsev$^{1}$}
\author{V.~M.~Shabaev$^{1}$}
\email[To whom all correspondence should be addressed: ]{v.shabaev@spbu.ru}
\author{R.~V.~Popov$^{1}$}
\author{Y.~S.~Kozhedub$^{1, 2}$}
\author{G.~Plunien $^{3}$}
\author{X.~Ma$^{4}$}
\author{Th.~St\"ohlker$^{5,6,7}$}
\author{D.~A.~Tumakov~$^{1}$}

\affiliation{
$^1$ Department of Physics, St. Petersburg State University,
Universitetskaya naberezhnaya 7/9, 199034 St. Petersburg, Russia\\
$^2$ NRC “Kurchatov Institute”, Academician Kurchatov 1, 123182 Moscow, Russia\\
$^3$ Institut f\"ur Theoretische Physik, Technische Universit\"at Dresden,
Mommsenstra{\ss}e 13, D-01062 Dresden, Germany\\
$^4$
Institute of Modern Physics, Chinese Academy of Sciences, 
730000  Lanzhou, China\\
$^5$
GSI Helmholtzzentrum f\"ur Schwerionenforschung GmbH,
Planckstrasse 1, D-64291 Darmstadt, Germany \\
$^6$Helmholtz-Institute Jena, D-07743 Jena, Germany\\
$^7$Theoretisch-Physikalisches Institut,
Friedrich-Schiller-Universit\"at Jena, D-07743 Jena, Germany
\vspace{10mm}
}
%
\begin{abstract}
In slow collisions of two bare nuclei with the total charge larger
than the critical value~$Z_{\rm cr} \approx 173$, the initially 
neutral vacuum can spontaneously decay into the charged vacuum 
and two positrons.
Detection of the spontaneous emission of positrons would be 
the direct evidence of this fundamental phenomenon. However, the spontaneously 
produced particles are indistinguishable from the dynamical background 
in the positron spectra. We show that the vacuum decay can 
nevertheless be observed 
via impact-sensitive measurements of pair-production probabilities. 
Possibility of such observation is demonstrated using numerical 
calculations of pair production
in low-energy collisions of heavy nuclei. 
\end{abstract}
\maketitle
In relativistic quantum mechanics, the energy levels of hydrogenlike 
ions are described by the Dirac equation. For the pointlike 
nucleus this equation has a solution for the 1$s$ state only if the nuclear
charge $Z$ is not greater 
than $Z_0 = 137$. 
Therefore the energy of this state 
is bounded from below by $E (Z_0) = 0$. However, for an extended nucleus 
$E(Z)$ decreases further as $Z$ increases and eventually crosses
the value $-mc^2$ at the critical 
charge~$Z_{\rm cr} \approx 173$~\cite{Pomeranchuk_45, Gershtein_70, 
Greiner_69, Zeldovich_71, Greiner_85, Mueller_94, Godunov_17}. 
After the crossing the level ``dives'' into the 
negative-energy Dirac continuum becoming 
a resonance. If this supercritical resonance state was 
initially vacant then it can be
occupied by two electrons from the negative-energy continuum 
with emission of two positrons~\cite{Gershtein_70, 
Greiner_69, Zeldovich_71, Greiner_85, Mueller_94}. 
This process can be interpreted as a spontaneous 
decay of the old neutral vacuum with formation of 
a new ``charged'' vacuum. 

Obviously, the required critical charge $Z_{\rm cr} \approx 173$ is much 
larger than the charge of the heaviest nuclei produced so far. 
However, two heavy colliding ions can form
a quasimolecular system with the total charge $Z_{\rm tot} = Z_1 + Z_2$
large enough for the ground state to reach the negative-energy continuum. 
Observation of the electron-positron pairs spontaneously produced during the 
collision would be the direct evidence of the vacuum decay. 
But in heavy-ion collisions the pair production is also induced by the ion dynamics. 
In order to detect  the vacuum decay, 
one has to distinguish the spontaneous pair production
from the dynamical one.

The experiments on low-energy heavy-ion collisions were intensively
performed many years ago at GSI (Darmstadt, Germany).
However, no sign of the spontaneous
pair production or the diving phenomenon had been found~\cite{Mueller_94, Ahmad_99}. 
There are several proposals for investigation of supercritical collisions at
the upcoming accelerator facilities~\cite{Gumbaridze_09, Ter-Akopian_15, Ma_17},
which will allow to perform the experiments on an entirely new level. 
In particular, experiments on low-energy collisions of heavy bare nuclei
are anticipated at these facilities. 
But so far it was not clear whether or not there exists a theoretical possibility 
of the diving phenomenon detection. 

To date, the pair production in low-energy ion collisions was investigated 
using various  theoretical approaches~\cite{Gershtein_73, Popov_73, Peitz_73, Lee_16,
Khriplovich_16_17, Reinhardt_81, Muller_88, 
Ackad_07_08, Maltsev_15, Bondarev_15, 
Maltsev_17, RV_Popov_18, Maltsev_18}. 
As was found by the Frankfurt group, the pair-production 
probability as a function of the total nuclear charge and the 
impact parameter has no threshold effects at the border of the supercritical 
region, where the spontaneous mechanism should start to work~\cite{Reinhardt_81}. 
It was also shown that the energy-differential 
spectra of the emitted positrons do not exhibit any feature
which can be associated with the spontaneous pair production. 
The calculations were performed using so-called monopole approximation,
in which only the spherical part of the two-center ion potential is taken into account. 
Recently the obtained results were confirmed 
with the monopole approximation~\cite{Maltsev_15}
as well as beyond it~\cite{Maltsev_17, RV_Popov_18, Maltsev_18}. 

The absence of any signature of the spontaneous mechanism
in the calculated pair-production probabilities and in the positron 
spectra led the Frankfurt group to the conclusion that 
the vacuum decay could only be observed in collisions with nuclear 
sticking, in which the nuclei are bound to each other for some period of time 
by nuclear forces~\cite{Soff_mem}. 
In such collisions, there should be a visible effect of the vacuum decay 
due to increase of the diving time.
In numerical calculations, the nuclear sticking
can be taken into account via introducing the time delay at the point 
of the closest nuclear approach. It was demonstrated that the 
time delay leads to the enhancement of 
the pair-production probability in the supercritical
case that can be explained only with 
the spontaneous mechanism (see, e.g., Ref.~\cite{Muller_88}).
However, to date there is no robust evidence of existence of 
the sufficiently long nuclear sticking. 

In this Letter, we show that the vacuum decay can be 
detected experimentally even without any nuclear sticking. 
The idea of the detection is based on the different 
behavior of the pair-production probability as a function of nuclear
velocities in the supercritical and subcritical cases. 

\begin{figure}
\centering 
\includegraphics[trim=0 0 0 0, clip, width = 0.7\textwidth]{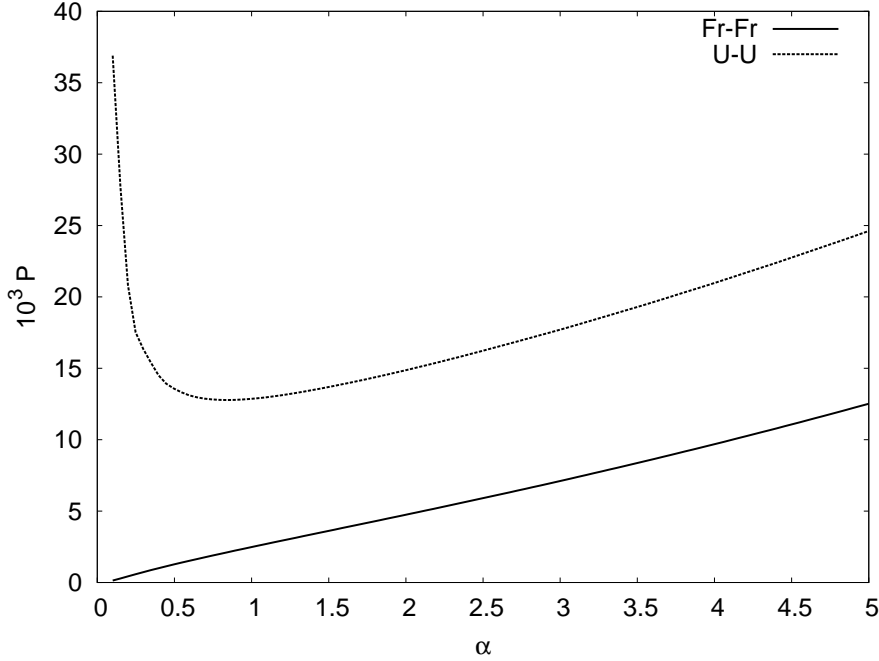}
\caption{Pair-production probability $P$ in the hypothetical head-on collision of bare nuclei 
with the modified dependence of the internuclear distance on time $R_\alpha (t)$,
defined by Eq.~(\ref{eq:art_trajectory}), 
as a function of $\alpha$. The solid line shows
the results for the Fr$-$Fr (subcritical) collision at $E=674.5$~MeV; 
the dashed line corresponds
to the U$-$U (supercritical) collision  at $E=740$~MeV.
The results were obtained in Ref.~\cite{Maltsev_15}.}
\label{fig:var_speed}
\end{figure}

Let us consider first hypothetical collisions with the modified
velocity~\cite{Maltsev_15}:
\begin{equation}
\dot{R}_\alpha (t)=\alpha \dot{R}(t).
\label{eq:art_trajectory}
\end{equation}
Here $R(t)$ is the internuclear distance which depends on 
time in accordance with the classical Rutherford scattering: 
\begin{equation}
 \begin{split}
  &R = a  \left( e \cosh \xi + 1 \right),\\
  &t = \sqrt{\frac{M_r a^3}{Z_1 Z_2}}  \left( e \sinh \xi + \xi \right),
 \end{split}
 \label{eq:ruth_traj}
\end{equation}
where
\begin{equation}
 a = \frac{Z_1 Z_2}{2 E}, 
 \qquad 
 e = \left( 1 + \frac{b^2}{a^2} \right)^{1/2},
 \qquad
 \xi \in (-\infty, \infty),
 \label{eq:a_e}
\end{equation}
$E$ is the collision energy in the center-of-mass frame, 
$M_r$ is the reduced mass of the nuclei,
and $b$ is the impact parameter. 
Varying the parameter~$\alpha$, we can change the nuclear velocity
in numerical calculations. Figure~\ref{fig:var_speed} presents the  
pair-production probability $P$ as a function of $\alpha$ obtained
in Ref.~\cite{Maltsev_15}. The calculations were performed 
for subcritical  Fr--Fr and supercritical U--U head-on collisions of bare 
nuclei at energy about the Coulomb barrier. As one can see from the 
figure, the behavior of the curves at small values of $\alpha$ is remarkably 
different. As $\alpha$ decreases, $P(\alpha)$ decreases in the subcritical
case and drastically increases in the supercritical one, which indicates the existence of 
the spontaneous pair production mechanism.  It should be emphasized that 
the subcritical curve {\it always} rises with increase of $\alpha$ 
and the supercritical curve has quite a simple shape with one minimum.

\begin{figure}
\centering 
\includegraphics[trim=0 0 0 0, clip, width = 0.7\textwidth]{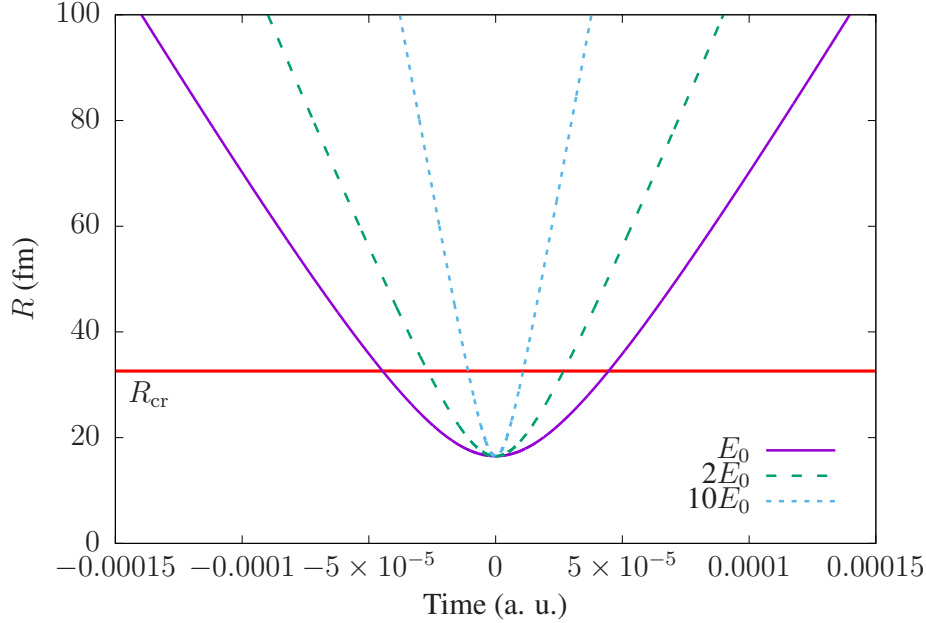}
\caption{The internuclear distance $R$ for U--U collision 
as a function of time for 
different values of the collision energy with the fixed distance of the closest approach
$R_{\rm min} =$ 16.5 fm, $E_0$ is the energy of the head-on collision.
The red horizontal line corresponds to the critical distance 
$R_{\rm cr} \approx 32.6$~fm 
and indicates the border between subcritical and supercritical regimes.}
\label{fig:traj}
\end{figure}

Of course, it is impossible to modify the collisions according 
to Eq.~\eqref{eq:art_trajectory} in real experiments. 
However, there exists a way to investigate the dependence of the 
pair-production probability on the ion velocity using the pure
Rutherford kinematics defined by Eq.~\eqref{eq:ruth_traj}. 
Let us fix the nuclear charges $Z_1$, $Z_2$ and  the distance 
of the closest nuclear
approach
\begin{equation}
  R_{\rm min} = a (e + 1).
\end{equation}
One can vary the collision energy~$E$ with changing
the impact parameter~$b$ according to the equation
\begin{equation}
  b^2 = R_{\rm min}^2 - R_{\rm min} \frac{Z_1 Z_2}{E}
  \label{eq:impact}
\end{equation}
with fixed $R_{\rm min}$.
The collision energy is bounded from below
by the value
\begin{equation}
  E_0 = \frac{Z_1 Z_2}{R_{\rm min}},
\end{equation}
which corresponds to the head-on collision ($b$ = 0). 
Using Eqs.~\eqref{eq:impact} and~\eqref{eq:ruth_traj},
for the range of available energies, $E \geq E_0$,
one can define the set of functions $R_E(t)$ which 
have the same minimum but different durations of the supercritical 
regime. For the case of U--U collision, these functions are displayed
in Fig.~\ref{fig:traj}. It can be seen that the supercritical time period
decreases monotonically with increase of $E$.
Employing the defined set of $R_E (t)$ it is possible to investigate
the pair-production probability as a function of nuclear velocity 
keeping the range of internuclear 
distances fixed ($R_{\rm min} \leq R(t) < \infty$).
The major limitation of this approach is that the nuclei cannot 
be slowed down more than it is allowed by the condition $E \geq E_0$.

In order to find the desired difference in pair production between subcritical and supercritical
systems, we performed calculations using the method described 
in Ref.~\cite{Maltsev_15}.  The method is based on the numerical solving of the 
time-dependent Dirac equation in the monopole approximation, according to 
which the two-center nuclear potential~$V_{\rm TC} (\vec r, t)$ is approximated 
by its spherically symmetric part
\begin{equation}
 V_{\rm mon} (r, t) = \frac{1}{4 \pi} \int d \Omega  \, V_{\rm TC} (\vec r, t).
 \label{eq:mon_pot}
\end{equation}
This approximation allows us to consider the radial Dirac equation instead of 
the two-center one. The corresponding electron wave function can be represented 
as 
\begin{equation}
 \psi_{\kappa m}(\vec{r},t) =
  \left  ( \begin{array}{l} \displaystyle
  \,\, \frac{~G_{\kappa}(r,t)}{r} \,  \chi_{\kappa m}(\Omega)
  \\[4mm] \displaystyle
  i \, \frac{F_{\kappa}(r,t)}{r} \, \chi_{-\kappa m}(\Omega)
  \end{array} \right ) \,,
  \label{eq:bispinor}
\end{equation}
where $\chi_{\pm \kappa m} (\Omega)$ are the spherical spinors, 
$F_{\kappa} (r, t)$ and $G_{\kappa} (r, t)$ are the small 
and large radial components, respectively, 
$m$ is the projection of the total angular momentum,
and $\kappa$ 
is the relativistic angular quantum number. We take into account
the electronic states with $\kappa = \pm 1$, which are expected 
to give the major contribution to the pair production. 
Since there is no coupling between these two sets of states,
the corresponding contributions can be calculated independently.

\begin{figure}[ht!]
\centering 
\includegraphics[trim=0 0 0 0, clip, width = 0.7\textwidth]{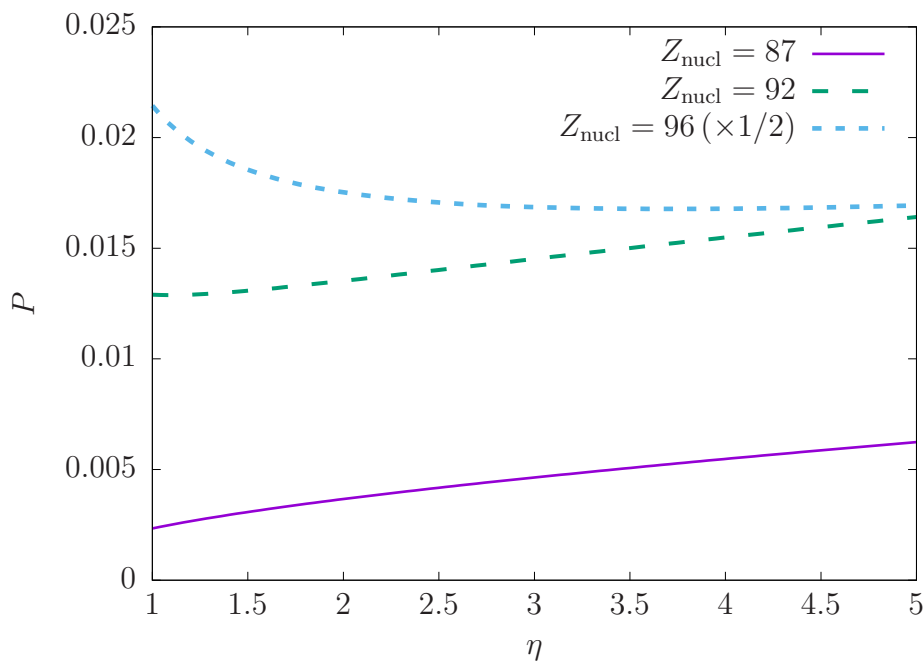}
\caption{The pair-production probability in the collision of two identical nuclei
with $Z_1 = Z_2 = Z_{\rm nucl}$
as a function of the ratio $\eta = E/E_0$, where 
$E$ is the collision energy and $E_0$ is the energy of the head-on collision. The 
results for $Z_{\rm nucl} = 96$ are multiplied by factor 0.5.}
\label{fig:p_e}
\end{figure}
For simplicity, we consider the collision of two identical bare nuclei
with $Z_1 = Z_2 = Z_{\rm nucl}$. The closest nuclear approach is
fixed to $R_{\rm min} =16.5$~fm. At such a distance the
nuclei are about 1--2~fm away from touching each other. 
The calculations are performed for subcritical and supercritical collisions 
at different energies~$E$ for different values of $Z_{\rm nucl}$. 
In Figure~\ref{fig:p_e}, we present the obtained results for the pair-production
probability $P$ as a function of the $\eta = E/E_0$ ratio for Fr--Fr ($Z_{\rm nucl} =$~87), 
U--U ($Z_{\rm nucl} =$~92), and Cm--Cm~($Z_{\rm nucl} =$~96)
collisions. The Fr--Fr system is subcritical (it is the heaviest subcritical system), 
the U--U and Cm--Cm systems are supercritical.
As can be seen from the figure,
the Fr--Fr curve goes monotonically down with decrease of $E$ as in the case of 
the modified collisions (see Fig.~\ref{fig:var_speed}). Such a behavior takes place 
for all collisions with $Z_{\rm nucl} \leq 87$. In contrast, the pair-production 
probability in the supercritical Cm--Cm collision starts to increase as 
$\eta$ approaches unity. In the U--U collision, which is also supercritical,
the function $P(\eta)$ has only a slight increase
as $\eta \rightarrow 1$ but exhibits clearly different behavior 
compared to the subcritical case.

\begin{figure}[h!]
\centering 
\includegraphics[trim=0 0 0 0, clip, width = 0.7\textwidth]{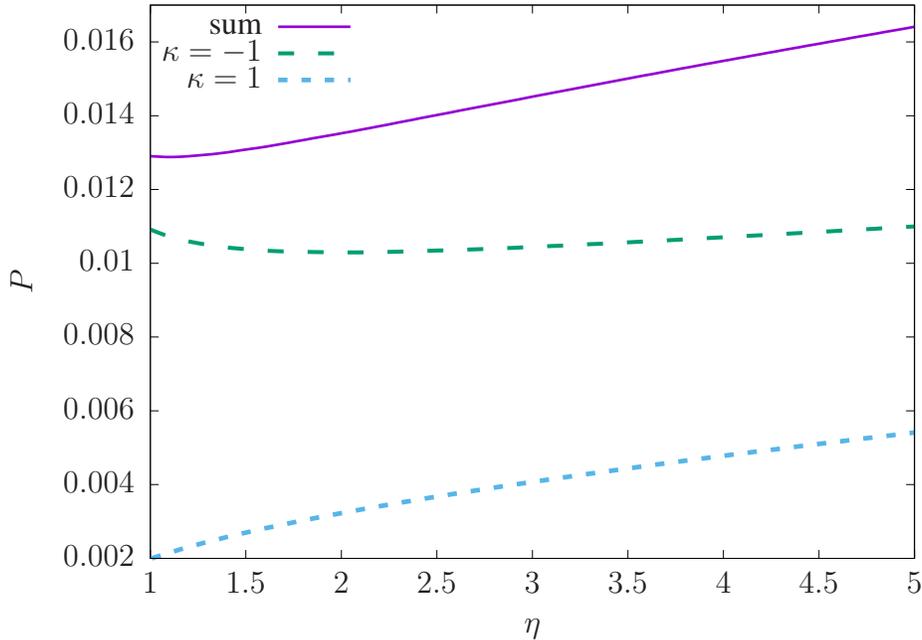}
\caption{The total pair-production probability in U--U collision, and contributions
from channels with $\kappa = \pm 1$ as functions of the ratio~$\eta = E/E_0$, where 
$E$ is the collision energy and $E_0$ is the energy of the head-on collision. }
\label{fig:u}
\end{figure}

To clarify this point, let us consider the U--U
collision in more details.
It should be noted that, in our calculations, the total pair-production probability 
is the sum of two independent contributions: 
$P_{\kappa = -1}$ and $P_{\kappa = 1}$, which correspond to creation of 
particles in the states with $\kappa = -1$ and $\kappa = 1$, respectively. 
Only the channel with $\kappa = -1$ is supercritical,  because it includes the 
diving 1$s$ state.  In Figure~\ref{fig:u}, we depict the calculated values
of $P_{\kappa = -1}$, $P_{\kappa = 1}$, and the total probability for the 
U--U collision. 
The curve corresponding to the supercritical ($\kappa = -1$) results 
has a rather pronounced
minimum while the subcritical ($\kappa = 1$) one is monotonic. But in 
the sum $P_{\kappa = -1} + P_{\kappa = 1}$
for $\eta \rightarrow 1$, an increase in $P_{\kappa = -1}$ and a 
decrease in $P_{\kappa = 1}$ almost cancel
each other out, which leads to a much less pronounced minimum.

\begin{figure}[h!]
\centering
\includegraphics[trim=0 0 0 0, clip, width = 0.7\textwidth]{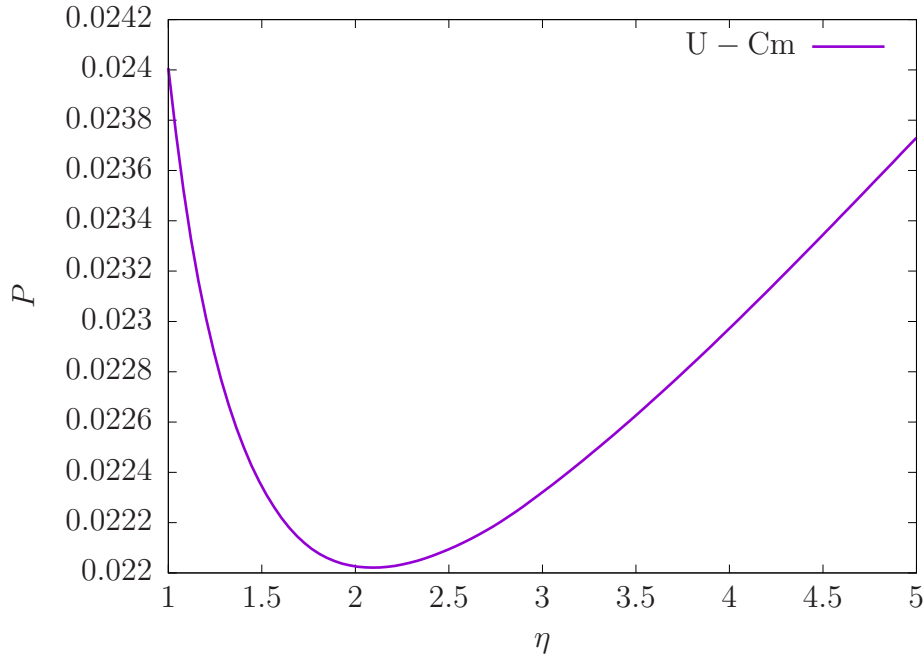}
\caption{ The total pair-production probability in U--Cm collision 
as a function of the ratio~$\eta = E/E_0$, where 
$E$ is the collision energy and $E_0$ is the energy of the head-on collision.}
\label{fig:u_cm}
\end{figure}

It is clear that all the calculations can be easily extended to 
asymmetric collisions of bare nuclei. 
In Figure~\ref{fig:u_cm}, we present the corresponding results 
for the pair-production probability as 
a function of $\eta$ for the U--Cm collision 
($Z_1 + Z_2 = 188$). 
As one can see, there is a clear signal of the spontaneous 
pair production for this system as well.

So far we considered only collisions of bare nuclei. 
On the one hand, the calculations for collisions of bare nuclei 
are the simplest to demonstrate the principle possibility 
of vacuum decay detection in the proposed scenario and, 
on the other hand, the experiments with such systems would be most 
favorable for such detection. 
Proposals for experimental investigation of 
collisions of bare nuclei up to Cm--Cm system were considered, 
e.g., in Ref.~\cite{Ter-Akopian_15}.  
However, we would like to note that the same scenario can be
potentially used for collisions of bare nuclei with atoms 
having the filled K-shell. 
An estimation of the bound quasimolecular level occupation
probability in collisions of bare uranium nuclei on uranium and
curium atoms with a filled K-shell, based on the methods developed
in Refs.~\cite{Tup_10, Tup_12, Kozhedub_14}, 
revealed that the filled K-shell can only lead to
decrease of the pair-production probability roughly by a factor
within the range 0.25 - 0.6. We also do not expect that a
possible energy dependence of the corresponding suppression
factor will qualitatively change the main conclusions concerning the
observation of the effect of interest. We note that even if such a
dependence is noticeable, it can be analyzed and isolated by means of
more accurate calculations of the effects of the filled K-shell.
The refined treatment requires more elaborate many-electron
two-center calculations of the pair-creation probabilities and
is currently under way.

\begin{figure}[h!]
\centering
\includegraphics[trim=0 0 0 0, clip, width = 0.7\textwidth]{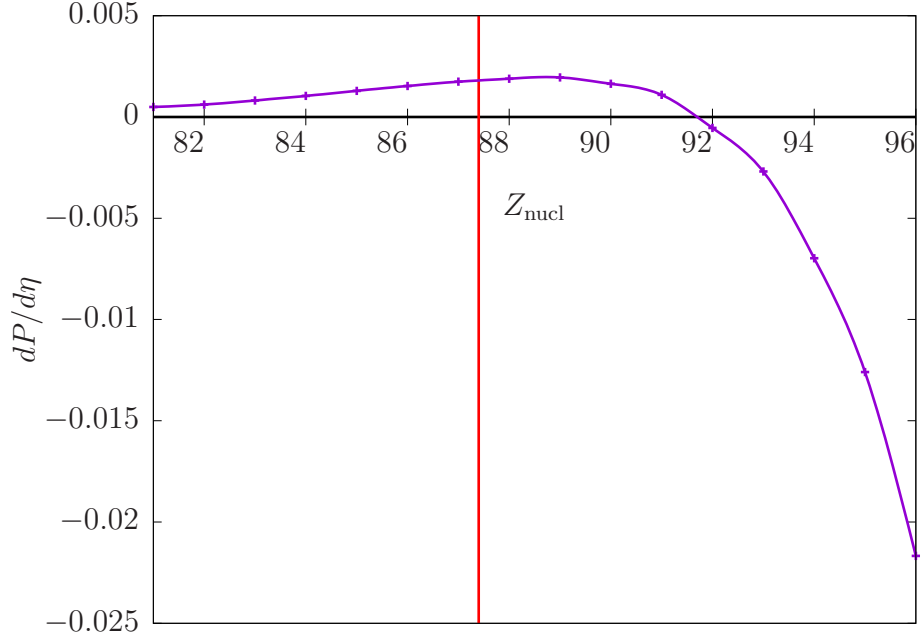}
\caption{ The derivative~$dP/d \eta$ taken at $\eta = 1$
as a function of $Z_{\rm nucl}$, where $\eta = E/E_0$,
$E$ is the collision energy, $E_0$ is the energy of the head-on collision, 
$P$ is the pair-production probability, and $Z_{\rm nucl}$ is the charge of each 
colliding nucleus. The red vertical line marks the border between subcritical 
and supercritical domains.}
\label{fig:der}
\end{figure}

In Figure~\ref{fig:der}, for symmetric collisions,
we show the derivative~$dP/d \eta$ taken at $\eta = 1$
as a function of $Z_{\rm nucl}$. As one can see from the figure, 
the function changes its behavior after transition to the supercritical
domain. It starts to decrease and finally crosses the zero line that 
corresponds to the appearance of the minimum on the graph of~$P(\eta)$.
The derivative becomes negative at $Z_{\rm nucl} \approx 92$.

From comparing the subcritical and supercritical scenarios, 
we conclude that there is the qualitative difference in 
behavior of the pair-production probability in the 
subcritical and the supercritical cases. 
If the distance of the closest approach is fixed,
the increase of this probability with decrease
of the collision energy can be observed only 
in the supercritical collisions. 
Moreover, even a pronounced decrease of $dP/d \eta$ at 
$\eta \approx 1$ as a function of $Z_{\rm nucl}$,
which takes place already at $Z_{\rm nucl} = 92$
(see Fig.~\ref{fig:der}), 
must be considered as a clear evidence of the vacuum decay 
at supercritical field. 

Although the calculations in the present paper are mainly restricted
to the monopole approximation, 
our recent study~\cite{Maltsev_17, RV_Popov_18, Maltsev_18} clearly
showed that effects beyond the  monopole approximation only
slightly change the pair-creation probabilities in the region of small
impact parameters, where the derivative presented in Fig.~\ref{fig:der} is 
calculated.
We can state that effects beyond the monopole approximation will not
change the main results obtained in this Letter. We believe, however,
that further studies of the pair-production probabilities and the
corresponding positron spectra beyond the monopole approximation can 
be very useful for finding the most promising experimental scenarios that 
allow for the determination of the angular distribution of the emitted 
positrons~\cite{Maltsev_18}.

We hope that the results obtained in this Letter will promote new efforts 
for the experimental detection 
of the vacuum decay in a supercritical Coulomb field.
In particular, such experiments seem feasible 
with the CRYRING facility at GSI/FAIR~\cite{Lestinsky_16, Hagmann_unub},
where storing of bare uranium nuclei at low energies is 
anticipated in the near future.

We thank I.~B.~Khriplovich and Yu.~Ts.~Oganessian for 
stimulating discussions. This work was supported by 
RFBR (Grants No.~18-32-00723 and No.~18-32-20063),
RFBR-NSFC (Grants No. 17-52-53136 and No. 11611530684),
and by SPbSU-DFG (Grants No.~11.65.41.2017 and No.~STO 346/5-1).
I.A.M. and R.V.P. 
acknowledge support from TU Dresden via the DAAD Programm Ostpartnerschaften.
The work of V.M.S. and R.V.P. was supported by the Foundation for 
the advancement of theoretical physics and mathematics ``BASIS''.
V.M.S. also acknowledges the support of the CAS President International 
Fellowship Initiative (PIFI) and of SPbSU (COLLAB 2019: No 37722582).
The research was carried out using computational resources provided by Resource 
Center ``Computer Center of SPbSU".

\end{document}